%% file: DESY-05-018.tex
\documentclass[zpreprint,zbstdefault]{zeus_paper}
\usepackage[english]{babel}
\input zeus_def.tex
\input  DESY-05-018-cit.tex
\begin{document}
\include{DESY-05-018-tit}
\include{DESY-05-018-aut}
\include{DESY-05-018-txt}
\include{DESY-05-018-ref}
\include{DESY-05-018-fig}
\end{document}

%% file: zeus_def.tex
\chardef\usc=95
\chardef\til=126
\catcode`\@=11 
\DeclareRobustCommand\xdotspace{\futurelet\@let@token\@xdotspace}
\def\@xdotspace{%
  \ifx\@let@token.\else
  \ifx\@let@token\bgroup.\else
  \ifx\@let@token\egroup.\else
  \ifx\@let@token\/.\else
  \ifx\@let@token\ .\else
  \ifx\@let@token~.\else
  \ifx\@let@token!.\else
  \ifx\@let@token,.\else
  \ifx\@let@token:.\else
  \ifx\@let@token;.\else
  \ifx\@let@token?.\else
  \ifx\@let@token/.\else
  \ifx\@let@token'.\else
  \ifx\@let@token).\else
  \ifx\@let@token-.\else
  \ifx\@let@token\@xobeysp.\else
  \ifx\@let@token\space.\else
  \ifx\@let@token\@sptoken.\else
   .\space
   \fi\fi\fi\fi\fi\fi\fi\fi\fi\fi\fi\fi\fi\fi\fi\fi\fi\fi}
\catcode`\@=12 

\newcommand{\stru}[2]{%
   \relax\ifmmode\hbox{\vrule height#1 depth#2 width0pt}%
   \else\vrule height#1 depth#2 width0pt\fi}

\newcommand{\Ronum}[1]{\uppercase\expandafter{\romannumeral#1}}
\newcommand{\ronum}[1]{\expandafter{\romannumeral#1}}
\DeclareRobustCommand{\LaTeXZ}{%
  \LaTeX\kern-.05em4\kern-.1em
  {\raisebox{-0.2ex}{$\scriptstyle\text{ZEUS}$}}\xspace}

\DeclareMathAlphabet{\mathbf}{OT1}{cmr}{bx}{sl}
\newcommand{\eVdist}{\kern-0.06667em}

\newcommand{\mev}{{\,\text{Me}\eVdist\text{V\/}}}
\newcommand{\gev}{{\,\text{Ge}\eVdist\text{V\/}}}

\newcommand{\cm}{\,\text{cm}}

\newcommand{\slashfrac}[2]{%
  \raisebox{0.5ex}{\ensuremath #1}\kern-0.12em/\kern-0.08em
  \raisebox{-.8ex}{\ensuremath #2}}

\newcommand{\sqr}[3]{%
    {\vcenter{\hrule height.#3ex\hbox{\vrule width.#2ex height#1ex
     \kern#1ex\vrule width.#3ex}\hrule height.#2ex}}}

\catcode`\@=11 
\newcommand{\parenbar}{\mathpalette\p@renb@r}
\def\p@renb@r#1#2{\vbox{%
  \ifx#1\scriptscriptstyle \dimen@.7em\dimen@ii.2em\else
  \ifx#1\scriptstyle \dimen@.8em\dimen@ii.25em\else
  \dimen@1em\dimen@ii.4em\fi\fi \offinterlineskip
  \ialign{\hfill##\hfill\cr
    \vbox{\hrule width\dimen@ii}\cr
    \noalign{\vskip-.3ex}%
    \hbox to\dimen@{$\mathchar300\hfil\mathchar301$}\cr
    \noalign{\vskip-.3ex}%
    $#1#2$\cr}}}
\catcode`\@=12 

\newcommand{\IP}{{\rm I$\kern-0.01667em$P}\xspace}

\mathchardef\qsm=63
\mathchardef\pls=43
\mathchardef\mns=512
\mathchardef\plm=518
\mathchardef\eql=61
\mathchardef\smallleft=300
\mathchardef\smallright=301
\mathchardef\les=316
\mathchardef\gre=318
\mathchardef\leq=532
\mathchardef\grq=533

\catcode`\@=11 
\newcounter{pict@width}
\newcounter{pict@height}
\newlength{\pict@scale}
\newcommand{\psfigadd}[4]{%
\setcounter{pict@width}{1*\ratio{#2+\pict@scale/2}{\pict@scale}}
\setcounter{pict@height}{1*\ratio{#3+\pict@scale/2}{\pict@scale}}
\setlength{\unitlength}{\pict@scale}
\hbox to #2{\hspace{-\fill}\begin{picture}(\thepict@width,\thepict@height)
\put(0,0){\psfig{figure=#1,width=#2,height=#3,clip=}}
\SetScale{0.283466457}
\SetWidth{1.763889}
{#4}
\end{picture}}
}
\newcounter{pict@widthfst}
\newcounter{pict@widthscd}
\newcounter{pict@widthtot}
\newcommand{\psfigaddtwo}[7]{%
\setcounter{pict@widthfst}{1*\ratio{#2+\pict@scale/2}{\pict@scale}}
\setcounter{pict@widthscd}{1*\ratio{#2+#4+\pict@scale/2}{\pict@scale}}
\setcounter{pict@widthtot}{1*\ratio{#2+#4+#6+\pict@scale/2}{\pict@scale}}
\setcounter{pict@height}{1*\ratio{#3+\pict@scale/2}{\pict@scale}}
\setlength{\unitlength}{\pict@scale}
\hbox{\hspace{-\fill}\begin{picture}(\thepict@widthtot,\thepict@height)
\put(0,0){\psfig{figure=#1,width=#2,height=#3,clip=}}
\put(\thepict@widthscd,0){\psfig{figure=#5,width=#6,height=#3,clip=}}
\SetScale{0.283466457}
\SetWidth{1.763889}
{#7}
\end{picture}}
}
\newcommand{\psfigror}[4]{%
\setcounter{pict@width}{1*\ratio{#2+\pict@scale/2}{\pict@scale}}
\setcounter{pict@height}{1*\ratio{#3+\pict@scale/2}{\pict@scale}}
\setlength{\unitlength}{\pict@scale}
\hbox{\begin{picture}(\thepict@width,\thepict@height)
\put(0,\thepict@height){\psfig{figure=#1,width=#3,height=#2,clip=,angle=270}}
\SetScale{0.283466457}
\SetWidth{1.763889}
{#4}
\end{picture}}
}
\newcommand{\psfigrol}[4]{%
\setcounter{pict@width}{1*\ratio{#2+\pict@scale/2}{\pict@scale}}
\setcounter{pict@height}{1*\ratio{#3+\pict@scale/2}{\pict@scale}}
\setlength{\unitlength}{\pict@scale}
\hbox{\begin{picture}(\thepict@width,\thepict@height)
\put(0,0){\psfig{figure=#1,width=#3,height=#2,clip=,angle=90}}
\SetScale{0.283466457}
\SetWidth{1.763889}
{#4}
\end{picture}}
}
\catcode`\@=12 
\newlength\listtextwidth

\catcode`\@=11 
\newlength{\@tabfninsert}
\newlength{\@tabfnwidth}
\newcommand{\tabfootnote}[2]{%
  \setlength{\@tabfninsert}{0.8em}
  \setlength{\@tabfnwidth}{\textwidth}
  \addtolength{\@tabfnwidth}{-\@tabfninsert}
  \addtolength{\@tabfnwidth}{-0.4em}
  \noindent\makebox[\@tabfninsert][r]{\footnotesize$^{#1}$\hfil}\hfill%
  \parbox[t]{\@tabfnwidth}{\footnotesize #2\hfill}}
\catcode`\@=12 

%% file: DESY-05-018-cit.tex
\def\citeCTD{{\cite{%
nim:a279:290,*npps:b32:181,*nim:a338:254%
}}\xspace}
\def\citeCAL{{\cite{%
nim:a309:77,*nim:a309:101,*nim:a321:356,*nim:a336:23%
}}\xspace}

%% file: DESY-05-018-tit.tex
\prepnum {DESY 05-018}

\title{
Search for pentaquarks decaying to $\Xi\pi$ in 
deep inelastic scattering at HERA
}                                                       
                    
\author{ZEUS Collaboration}

\date{January, 2005}


\abstract{
A search for pentaquarks decaying to $\Xi^{-}\pi^{-}$ ($\Xi^{-}\pi^{+}$) 
and corresponding antiparticles has been performed with the ZEUS detector
at HERA. The data sample consists of deep inelastic $ep$ scattering events
at centre-of-mass energies of 300 and 318 $\gev$, and corresponds to 
$121$~pb$^{-1}$ of
integrated luminosity. 
A clear signal for $\Xi^{0}(1530)\rightarrow\Xi^{-}\pi^{+}$ was observed.
However, no signal for any new baryonic state was observed at 
higher masses in either  
the $\Xi^{-}\pi^{-}$ or  $\Xi^{-}\pi^{+}$ 
channels. The searches in the antiparticle
channels were also negative.
Upper limits on the ratio of a possible $\Xi^{--}_{3/2}$ ($\Xi^{0}_{3/2}$) 
signal to the $\Xi^{0}(1530)$ signal 
were set in the mass range 1650--2350 $\mev$.
}

\makezeustitle

%% file: DESY-05-018-aut.tex
\pagenumbering{Roman}                                                                              
                                                   %
\begin{center}                                                                                     
{                      \Large  The ZEUS Collaboration              }                               
\end{center}                                                                                       
  S.~Chekanov,                                                                                     
  M.~Derrick,                                                                                      
  S.~Magill,                                                                                       
  S.~Miglioranzi$^{   1}$,                                                                         
  B.~Musgrave,                                                                                     
  \mbox{J.~Repond},                                                                                
  R.~Yoshida\\                                                                                     
 {\it Argonne National Laboratory, Argonne, Illinois 60439-4815}, USA~$^{n}$                       
\par \filbreak                                                                                     
  M.C.K.~Mattingly \\                                                                              
 {\it Andrews University, Berrien Springs, Michigan 49104-0380}, USA                               
\par \filbreak                                                                                     
  N.~Pavel, A.G.~Yag\"ues Molina \\                                                                
  {\it Institut f\"ur Physik der Humboldt-Universit\"at zu Berlin,                                 
           Berlin, Germany}
% ...
\newpage                                                                                           
$^{\    1}$ also affiliated with University College London, UK \\                                  
% ...
                                                           %
                                                           %
\newpage   
                                                           %
                                                           %
\begin{tabular}[h]{rp{14cm}}                                                                       
$^{a}$ &  supported by the Natural Sciences and Engineering Research Council of Canada (NSERC) \\  
$^{b}$ &  supported by the German Federal Ministry for Education and Research (BMBF), under        
          contract numbers HZ1GUA 2, HZ1GUB 0, HZ1PDA 5, HZ1VFA 5\\                                
$^{c}$ &  supported in part by the MINERVA Gesellschaft f\"ur Forschung GmbH, the Israel Science   
          Foundation (grant no. 293/02-11.2), the U.S.-Israel Binational Science Foundation and    
          the Benozyio Center for High Energy Physics\\                                            
$^{d}$ &  supported by the German-Israeli Foundation and the Israel Science Foundation\\           
$^{e}$ &  supported by the Italian National Institute for Nuclear Physics (INFN) \\                
$^{f}$ &  supported by the Japanese Ministry of Education, Culture, Sports, Science and Technology 
          (MEXT) and its grants for Scientific Research\\                                          
$^{g}$ &  supported by the Korean Ministry of Education and Korea Science and Engineering          
          Foundation\\                                                                             
$^{h}$ &  supported by the Netherlands Foundation for Research on Matter (FOM)\\                   
$^{i}$ &  supported by the Polish State Committee for Scientific Research, grant no.               
          620/E-77/SPB/DESY/P-03/DZ 117/2003-2005 and grant no. 1P03B07427/2004-2006\\             
$^{j}$ &  partially supported by the German Federal Ministry for Education and Research (BMBF)\\   
$^{k}$ &  supported by RF Presidential grant N 1685.2003.2 for the leading scientific schools and  
          by the Russian Ministry of Education and Science through its grant for Scientific        
          Research on High Energy Physics\\                                                        
$^{l}$ &  supported by the Spanish Ministry of Education and Science through funds provided by     
          CICYT\\                                                                                  
$^{m}$ &  supported by the Particle Physics and Astronomy Research Council, UK\\                   
$^{n}$ &  supported by the US Department of Energy\\                                               
$^{o}$ &  supported by the US National Science Foundation\\                                        
$^{p}$ &  supported by the Polish Ministry of Scientific Research and Information Technology,      
          grant no. 112/E-356/SPUB/DESY/P-03/DZ 116/2003-2005 and 1 P03B 065 27\\                  
$^{q}$ &  supported by FNRS and its associated funds (IISN and FRIA) and by an Inter-University    
          Attraction Poles Programme subsidised by the Belgian Federal Science Policy Office\\     
\end{tabular}                                                                                      
                                                           %
                                                           %

%% file: DESY-05-018-txt.tex
\pagenumbering{arabic} 
\pagestyle{plain}
\section{Introduction}

A number of experiments 
\cite{prl91012003,*plb572:127,*prl91252001,*prl92:032001,*pan66:1715,*pan67:682,*hep-ex-0401024,*plb585:213,*hep-ex-0403011,*hep-ex/0410016} 
including ZEUS \cite{plb591:0722} 
have reported narrow signals in the vicinity of $1530\mev$ 
in the $nK^{+}$ and $pK^{0}_{S}$ invariant-mass spectra.  
The signals are consistent with the exotic pentaquark baryon state $\Theta^{+}$
with quark content $uudd\bar{s}$ \cite{zp:a359:305}.
Several other experiments have searched for this state with negative results
\cite{prd:70:012004,prl:93:212003,ejp:20:455,*jphys30s1201,pl:599:1}.

The $\Theta^+$ lies at the apex of a hypothetical antidecuplet of pentaquarks
with spin $1/2$ \cite{zp:a359:305}. The baryonic states
$\Xi^{--}_{3/2}$ and $\Xi^{+}_{3/2}$ at the bottom of this
antidecuplet  are also manifestly exotic. 
According to Diakonov et al. \cite{zp:a359:305},  the
members of the antidecuplet, which belong to the isospin
quartet of $S=-2$ baryons, have a mass of
about $2070\mev$ and a partial decay width into $\Xi\pi$ of about
$40\mev$. On the other hand, Jaffe and Wilczek \cite{prl:91:232003} predicted 
a mass around $1750\mev$ and a width $50\%$ 
larger for these states than that of the $\Theta^{+}$. 
The isospin $3/2$ multiplet contains two states
with ordinary charge assignments $(\Xi^{0}_{3/2},\Xi^{-}_{3/2})$
in addition to the exotic states
$\Xi^{+}_{3/2}(uuss\bar{d})$ and $\Xi^{--}_{3/2}(ddss\bar{u})$.
Recently, NA49 \cite{prl92:042003} at the CERN SPS  
reported the observation 
of the $\Xi^{--}_{3/2}$ and $\Xi^{0}_{3/2}$ members of the 
$\Xi_{3/2}$ multiplet, with a mass of $1862\pm 2\mev$ and 
a width below $18\mev$. 
The signals were also seen in the corresponding antibaryon spectra. 
However, searches for such resonances by other experiments  
\cite{hep-ex-0405042,pl:599:1,prd:70:012004,prl:93:212003,hep-ex/0412027}
were negative.

This paper describes a search for new baryonic states 
in the $\Xi^{-}\pi^{\pm}$ and $\bar{\Xi}^{+}\pi^{\pm}$ invariant-mass 
spectra in $ep$ collisions measured with the ZEUS detector at HERA.
The studies were  performed in the central pseudorapidity region
where hadron production is dominated by fragmentation. 
The analysis was restricted to the 
deep inelastic scattering (DIS)
regime, and the $\Xi^{-}$($\bar{\Xi}^{+}$) states were reconstructed 
via the $\Lambda\pi^{-}$($\bar{\Lambda}\pi^{+}$) decay channel.

\section{Experimental setup}

ZEUS is a multipurpose detector described in detail
elsewhere \cite{zeus:1993:bluebook}.
The main components used in the
present study  are the central tracking detector and
the uranium-scintillator calorimeter.

The central tracking detector (CTD) \citeCTD is  a cylindrical
drift chamber with nine superlayers covering the  
polar-angle\footnote{The ZEUS coordinate 
system is a right-handed Cartesian system, with
the $Z$ axis pointing in the proton beam direction, referred to
as the ``forward direction'', and the $X$ axis
pointing left toward the center of HERA. The coordinate origin is at the
nominal interaction point.}
region $15^o \le  \theta \le  164^o$ and the
radial range 18.2--79.4 cm. 
The transverse-momentum resolution
for charged tracks traversing all CTD layers is
$\sigma(p_{T})/p_{T} = 0.0058 p_{T}
\oplus 0.0065 \oplus  0.0014/p_{T}$, 
with $p_{T}$ in $\gev$. To estimate the energy loss per unit length, $dE/dx$,
of particles in the CTD \cite{pl:b481:213,*epj:c18:625}, the truncated mean 
of the anode-wire pulse heights was calculated, which removes the lowest $10\%$
and at least the highest $30\%$ depending on the number of saturated hits.
The measured $dE/dx$ values were normalised to the $dE/dx$ peak
position for tracks with momenta $0.3<p<0.4 \gev$, 
the region of minimum ionisation for pions. 
Henceforth, $dE/dx$ is quoted in units of minimum ionising
particles (mips). The resolution of the $dE/dx$ measurement for full-length
tracks is about $9\%$. The tracking system was used to establish the 
primary and secondary vertices.

The CTD
is  surrounded by the uranium-scintillator 
calorimeter, the CAL \citeCAL, which is divided
into three parts: forward, barrel and rear.
The calorimeter is longitudinally segmented into electromagnetic
and hadronic sections. The smallest subdivision of the calorimeter 
is called a cell. 
The energy resolution of the calorimeter
under test-beam conditions
is $\sigma_E/E=0.18/\sqrt{E}$ for electrons and
$\sigma_E/E=0.35/\sqrt{E}$ for hadrons (with $E$ in $\gev$).
A presampler \cite{nim:a382:419,*magill:bpre} mounted in front of the 
calorimeter was used to correct the energy of the scattered 
electron\footnote{From now on, the word ``electron''
is used as a generic term for either electrons or positrons.}.
The position of electrons scattered with a small angle was measured using 
the small-angle rear tracking 
detector (SRTD) \cite{nim:a401:63}.

The luminosity was determined from the rate of the bremsstrahlung process 
$ep \to ep \gamma$, where the photon was measured with a lead-scintillator
calorimeter \cite{Desy-92-066,*zfp:c63:391,*acpp:b32:2025} located at 
$Z=-107$ m.
\section{Data sample}

The data sample for this analysis was taken 
during the 1996--2000 running period of HERA, 
and corresponds to an integrated luminosity of
$121$~pb$^{-1}$. 
The electron-beam energy was $27.5\gev$ and the proton-beam
energy was $820\gev$ for the 96--97 running period and $920\gev$ for the 98--00
running period. 

The exchanged photon virtuality, $Q^2$, 
was reconstructed from the energy and angle of the scattered electron.
The scattered-electron candidate was
identified from the pattern  of
energy deposits in the CAL \cite{nim:a365:508}. 
The following requirements were used to select neutral current DIS events:

\begin{itemize}

\item[$\bullet$]
$E_{e^{'}}\geq 5\gev$, where $E_{e^{'}}$ is the
energy of the scattered electron;

\item[$\bullet$]
a primary vertex position in the range
$\mid Z_{\mathrm{vertex}} \mid \le 50$ cm;

\item[$\bullet$]
$35\> \leq\> \sum E_i(1-\cos\theta_i)  \>\leq\> 60\gev$, where
$E_i$ is the energy of the $i$th calorimeter
cell and $\theta_i$ is its polar angle with respect to the measured primary
vertex position, 
and the sum runs over all cells;

\item[$\bullet$]
$Q^{2} > 1\gev^{2}$.

\end{itemize}

The present analysis was based on tracks measured in the CTD.
All tracks were required to pass through at least three CTD superlayers.
This requirement corresponds to the 
pseudorapidity range $|\eta| < 1.75$ in 
the laboratory frame.
Only tracks with transverse momenta 
$p_{T}^{\mathrm{lab}} > 150\mev$ were considered. The above cuts restricted
this analysis to a region where the track acceptance 
and resolution of the CTD are high.

The energy-loss measurement 
in the CTD, $dE/dx$, was used for particle identification.
Tracks with $f < dE/dx < F$ were taken as (anti)proton candidates, 
where $f=0.3/p^{2}+0.8$ and $F=1.0/p^{2}+1.2$ 
($p$ is the total track momentum in $\gev$) are functions 
motivated by the Bethe-Bloch equation.
The $dE/dx$ requirements for $\pi^{+}$($\pi^{-}$) candidates were 
$dE/dx<((0.1/p^{2})+0.8)$ or $dE/dx<1.8$~mips.

Candidates for long-lived neutral strange hadrons decaying to two charged 
particles were  identified by selecting pairs of oppositely charged tracks
fitted to a displaced secondary vertex. Events were required to have
at least one such candidate.

As a first step in the $\Xi^{-}\pi^{-}(\Xi^{-}\pi^{+})$ 
invariant-mass reconstruction, $\Lambda$ baryons were identified.  
Then, the $\Lambda$ baryons were combined with a $\pi^{-}$ to 
form $\Xi^{-}$ candidates. 
Finally, the 
$\Xi^{-}\pi^{-}(\Xi^{-}\pi^{+})$ invariant mass was reconstructed using 
pions associated with the primary vertex. 
The same procedure was used for the antiparticles.

The $\Lambda$ baryons  were identified by their
charged decay mode,  $\Lambda\to p\pi^{-}$, using pairs of tracks 
from secondary vertices. 
In order to reduce background further,
the track with the higher momentum was required to have a $dE/dx$ consistent 
with that of a proton and was assigned the proton mass.
The resulting invariant-mass spectra
for $p\pi^{-}$ and $\bar{p}\pi^{+}$  are  shown in 
Fig.~\ref{lam_mass_spec}. 
The measured number of $\Lambda$ baryons, as well as the
background under the peak, are higher for the $p\pi^{-}$ than
for the  $\bar{p}\pi^{+}$  spectrum. This is because of tracks produced 
in secondary interactions in the beampipe.    

To reconstruct $\Xi^{-}$ candidates, $\Lambda$ candidates with invariant masses
in the range 1111--1121$\mev$ were combined with negatively charged tracks. 
The distance of 
closest approach (DCA) in three dimensions between the trajectories of 
the $\Lambda$ and the $\pi^{-}$ was calculated and a
cut on the DCA of $1.0\cm$ was used to select preferentially those coming
from the same vertex. In addition, the following  
cuts were applied to increase the
significance of the $\Xi^{-}$ signal: 

\begin {itemize}

\item[$\bullet$]
the distance between the decay of the $\Xi^{-}$ and 
the primary vertex was required to be larger than
$1.75\cm$, since most of the
combinatorial background comes from tracks originating from the primary vertex
\cite{thesis:ziegler:2002}; 

\item[$\bullet$]
the momentum of the $\pi^{-}$ candidate was required to be less 
than the momentum of the 
$\Lambda$ candidate,   
since in the $\Xi^{-}$ decay, 
the $\Lambda$ takes the largest fraction of the $\Xi^{-}$ momentum; 

\item[$\bullet$]
the $\Lambda$ decay was required to be further from the primary vertex
than the $\Xi^{-}$ decay.
 
\end {itemize}

The resulting 
$\Lambda \pi^{-}$ and $\bar{\Lambda}\pi^{+}$ invariant-mass spectra are 
shown in Fig.~\ref{xi_mass_spec}. Because of the short decay length of the 
$\Xi$,
the cuts eliminate the contributions from secondary interactions in the 
beampipe; the number of $\Xi^{-}$ and $\bar{\Xi}^{+}$ are the same within
their uncertainties.
The measured width of the $\bar{\Xi}^{+}$ is somewhat larger than that of the
$\Xi^{-}$ as expected from the different momentum resolution for 
positive and negative particles.
The $\Xi^{-}$ candidates were selected within the invariant-mass 
range 1317--1327~$\mev$, shown as the shaded areas.  
In order to decrease the combinatorial background, only events 
with one $\Xi^{-}$ candidate, which comprises $92\%$ of the whole 
sample, were retained.

To search for the exotic $\Xi^{--}_{3/2}$ state, 
the selected $\Xi^{-}$ candidates 
were combined with $\pi^{-}$ tracks from the primary vertex. 
To reduce background, only tracks with momenta 
smaller than those of the $\Xi^{-}$ were used.
Analogous searches were performed for the $\bar{\Xi}^{++}_{3/2}$, 
$\Xi^{0}_{3/2}$ and $\bar{\Xi}^{0}_{3/2}$.

A number of checks were carried out to verify the
robustness of the above reconstruction procedure: 
(a) the $dE/dx$ requirements for pions and protons were removed;
(b) events with multiple  $\Xi^{-}$ candidates 
were retained;  
(c) the cut on the DCA was varied between $0.5\cm$ and $3.0\cm$; 
(d) the cut on the distance between the decay position 
of the $\Xi^-$ and the primary vertex was
varied between $1.0\cm$ and $3.0\cm$.   
In all these cases, 
the variations had a small impact on the reconstruction efficiency of  
the $\Xi^-\to \Lambda\pi^-$ and $\Xi^{0}(1530) \to \Xi^{-}\pi^{+}$ 
decay channels, or 
led to a reduction of the signal-to-background ratio at 
the level of 10--30$\%$.

\section{Results and Conclusions}

The resulting $\Xi\pi$ invariant-mass spectrum for the sum of
all four charge combinations, 
$\Xi^{-}\pi^{-}$, $\Xi^{-}\pi^{+}$, $\bar{\Xi}^{+}\pi^{-}$, 
$\bar{\Xi}^{+}\pi^{+}$ 
is shown in Fig.~\ref{penta_mass_spec}a for $Q^2>1\gev^2$.
The invariant-mass spectra for each $\Xi\pi$ combination separately   
are shown in Fig.~\ref{penta_mass_spec_each}.
In both the $\Xi^{-}\pi^{+}$ and $\bar{\Xi}^{+}\pi^{-}$ spectra, 
the well established $\Xi^{0}(1530)$ state \cite{pl:b592:2} is observed. 
The $\Xi^{0}(1530)$ peak was fitted by a Gaussian, and 
the background was parametrised by the function:  
\begin{equation}
\label{bgeqn}
B(M)=P_{1}(M-m_{\Xi}-m_{\pi})^{P_{2}} e^{-P_{3}(M-m_{\Xi}-m_{\pi})},
\end{equation}
where $M$ is the $\Xi\pi$ invariant mass, $m_{\Xi}$ and 
$m_{\pi}$ are the masses
of the $\Xi$ and the $\pi$, respectively, and $P_{1}$, $P_{2}$ and 
$P_{3}$ are free parameters. 
The extracted number 
of signal events was $192\pm30$. 
The measured peak position of $1533.3\pm 1.0\> \mathrm{(stat.)} \mev$ 
is consistent with the PDG value
\cite{pl:b592:2}, taking into
account a systematic uncertainty of 1--2$\mev$ on the mass 
measurement. The measured width of $6.6\pm1.4\> \mathrm{(stat.)}\mev$ is 
consistent with the detector resolution.
No signal is observed 
near $1860\mev$ in any of the spectra.

A similar analysis was performed for $Q^{2}>20\gev^2$, a kinematic region
of DIS where the $\Theta^{+}$ state was most clearly observed
by ZEUS \cite{plb591:0722}. 
The resulting invariant-mass
spectrum of $\Xi\pi$ for the sum of all four charge combinations
is shown in Fig.~\ref{penta_mass_spec}b. 
Again, no signal is observed near $1862~\mev$.  

In addition to the peak near $1530\mev$ 
due to the established $\Xi^{0}(1530)$ state, a possible peak
near $1690\mev$ for $Q^2>20\gev^2$ is observed (Fig.~\ref{penta_mass_spec}b).  
This enhancement could be due to the $\Xi(1690)$ baryon, 
the properties of which are not 
well determined \cite{pl:b592:2}.
Using the fit described in Eq.~(\ref{bgeqn}), 
the mass of this peak was found to 
be at $1687.5\pm 4.0\> \mathrm{(stat.)}\mev$,
and the Gaussian width was $9.5\pm 3.7\> \mathrm{(stat.)} \mev$.   
The statistical significance of this signal is $2.5 \sigma$. 

According to the NA49 study \cite{prl92:042003}, two additional cuts
were used to reduce the background for the pentaquark searches:  
a cut on the opening angle between the $\Xi$ and the $\pi$, 
and a cut on the momentum of the pion used in the reconstruction of the
$\Xi\pi$ invariant mass.
These two cuts were also tried in this analysis, with
the opening-angle cut varied from $0.1$ to $0.5$ radians, 
and the $\pi$ momentum
cut varied from $0.5 \gev$ to $1.5 \gev$. 
In no case was a pentaquark signal seen.
However, these two cuts changed the background shape by rejecting 
events near the $\Xi\pi$ mass threshold, reducing or completely suppressing the
$\Xi^{0}(1530)$ signal.

Given the absence of a signal in our data, 
$95\%$ C.L. limits were set on the production of new states
decaying to $\Xi^-\pi^-$ or $\Xi^-\pi^+$ in the mass range 1650--2350 $\mev$
in DIS for $Q^2>1\gev^2$. 
The upper limits on the ratio, $R$, of the number of
events in a (sliding) mass window to the reconstructed 
number of $\Xi^0(1530)$ events are presented.
This ratio is a good indicator of the sensitivity to a new state, given the
robust signal observed for the established $\Xi^{0}(1530)$ 
state. Most of the acceptances and reconstruction 
efficiencies largely cancelled 
in the ratio; some residual effects are present
since the acceptance has a dependence on the mass of the state. 
For example, the rapidity distribution 
of the $\Xi\pi$ combinations in the centre-of-mass system 
changes markedly over the mass range. 

The limits were set using Bayesian statistics assuming flat prior distributions
for $R$. The width of the search 
window was set equal
to $26.4\mev$, which is $\pm 2\sigma$ of the measured width of the 
$\Xi^{0}(1530)$. The background was modelled using Eq.~(\ref{bgeqn}).
 The $95\%$ C.L. limits on $R$ varied between
$0.1$ to $0.5$ as a function of the central value of 
the mass window, as shown in
Fig.~\ref{penta_mass_spec}a. In the NA49
signal region, $R$ is less than 0.29 at the $95\%$ C.L.

In addition to the above method, the  $95\%$ C.L. limits on the ratio $R$  
were calculated assuming a Gaussian probability function in
the unified approach \cite{pr:d57:3873} and fixing the reconstruction width
of the expected pentaquark state to $10\mev$, which is close to
expectations. Similar values for the limits were obtained.  

The number of $\Xi^{0}(1530)$ signal events reconstructed in this analysis is 
about the same as for the NA49 data analysed 
without the opening-angle and momentum 
cuts \cite{jphysg30s1359}. Therefore, 
the statistical sensitivity should be about the same
for the two analyses in this mass region. However, 
it should be
noted that NA49 is a fixed target experiment, which has good acceptance in the
forward region. The non-observation of this signal 
in the central-fragmentation region
in the ZEUS data does not necessarily contradict the 
observation of a signal predominantly 
produced in the forward region. 

In conclusion, a search for new baryons that decay to $\Xi^{-}\pi^{-}$ 
and $\Xi^{-}\pi^{+}$ was performed with the ZEUS detector using a DIS 
data sample with $Q^{2}>1\gev^{2}$ 
corresponding to an integrated luminosity of $121$~pb$^{-1}$. 
No pentaquark signal was found. Upper limits at $95\%$ C.L. on the ratio of 
the 
$\Xi^{--}_{3/2}$($\Xi^{0}_{3/2}$) signal to the $\Xi^{0}(1530)$ are
set in the mass range 1650--2350$\mev$.

\section*{Acknowledgments}
\vspace{0.3cm}
We thank the DESY Directorate for their strong support and encouragement.
The remarkable achievements of the HERA machine group were essential for
the successful completion of this work.
The design, construction and installation of the ZEUS detector have been
made possible by the effort of many people who are not listed as authors.

\vfill\eject

%% file: DESY-05-018-ref.tex
{
\def\bibname{\Large\bf References}
\def\refname{\Large\bf References}
\pagestyle{plain}
\ifzeusbst
  \bibliographystyle{./BiBTeX/bst/l4z_default}
\fi
\ifzdrftbst
  \bibliographystyle{./BiBTeX/bst/l4z_draft}
\fi
\ifzbstepj
  \bibliographystyle{./BiBTeX/bst/l4z_epj}
\fi
\ifzbstnp
  \bibliographystyle{./BiBTeX/bst/l4z_np}
\fi
\ifzbstpl
  \bibliographystyle{./BiBTeX/bst/l4z_pl}
\fi
{\raggedright
\bibliography{./BiBTeX/user/syn.bib,%
              ./BiBTeX/bib/l4z_articles.bib,%
              ./BiBTeX/bib/l4z_books.bib,%
              ./BiBTeX/bib/l4z_conferences.bib,%
              ./BiBTeX/bib/l4z_h1.bib,%
              ./BiBTeX/bib/l4z_misc.bib,%
              ./BiBTeX/bib/l4z_old.bib,%
              ./BiBTeX/bib/l4z_preprints.bib,%
              ./BiBTeX/bib/l4z_replaced.bib,%
              ./BiBTeX/bib/l4z_temporary.bib,%
              ./BiBTeX/bib/l4z_zeus.bib}}
}
\vfill\eject

%% file: DESY-05-018-fig.tex

\begin{figure}
\begin{center}
\includegraphics[width=16.0cm]{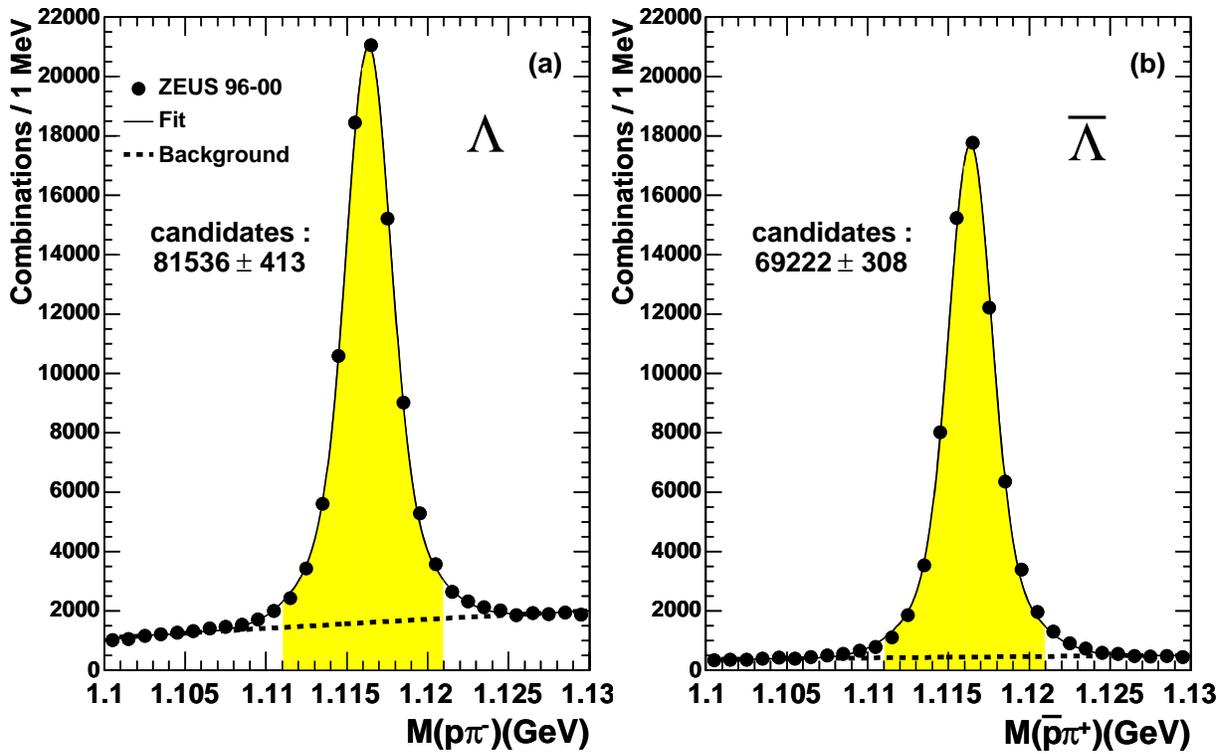}%
\caption{The (a) $p\pi^{-}$ and (b) $\bar{p}\pi^{+}$ invariant-mass spectra for 
$Q^{2}>1\gev^{2}$. The solid line shows the result of a fit 
using a double Gaussian plus a first-order polynomial  
background function, while the dashed line shows the background. 
The shaded areas indicate the mass range of the selected candidates.} 
\label{lam_mass_spec}
\end{center}
\end{figure}

\newpage
\begin{figure}
\begin{center}
\includegraphics[width=16.0cm]{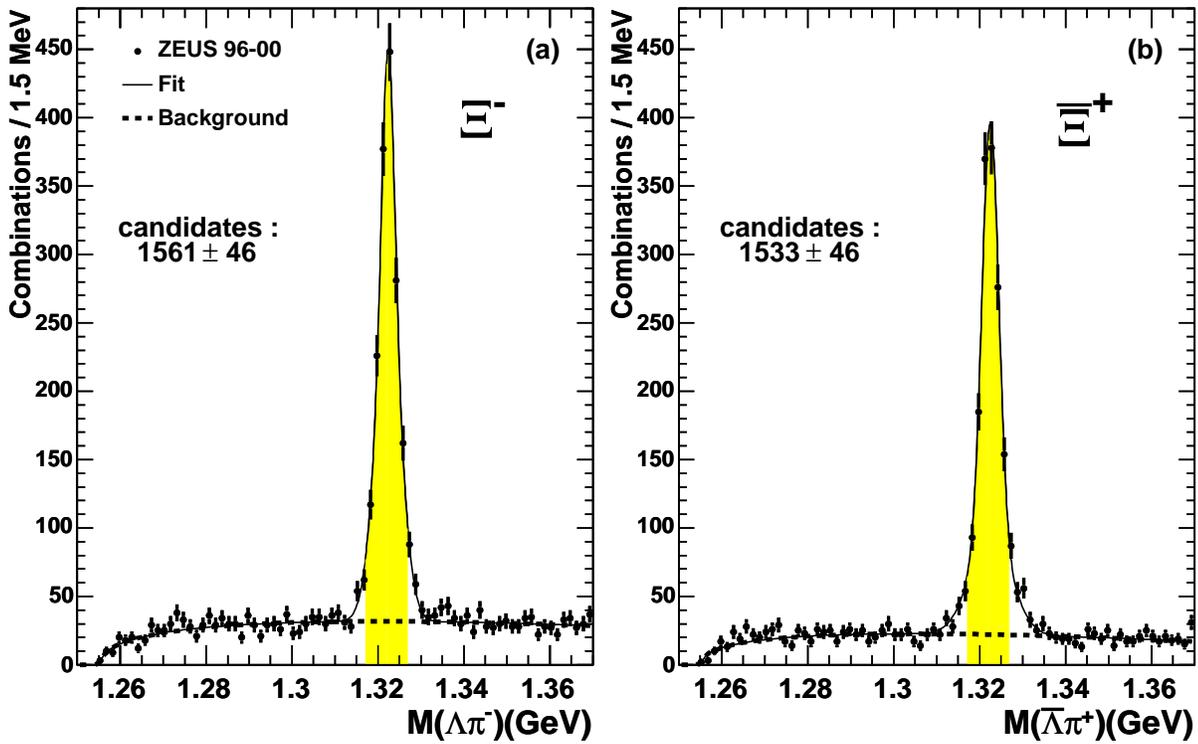}%
\caption{The (a) $\Lambda\pi^{-}$ and (b) $\bar{\Lambda}\pi^{+}$ 
invariant-mass 
spectra for $Q^{2}>1\gev^{2}$.
The solid line shows the result of a fit 
using a double Gaussian plus a 
second-order polynomial background function, 
while the dashed line shows the background.
The shaded areas show the mass range of the selected candidates.
} 
\label{xi_mass_spec}
\end{center}
\end{figure}


\newpage
\begin{figure}
\begin{center}
\includegraphics[width=16.0cm]{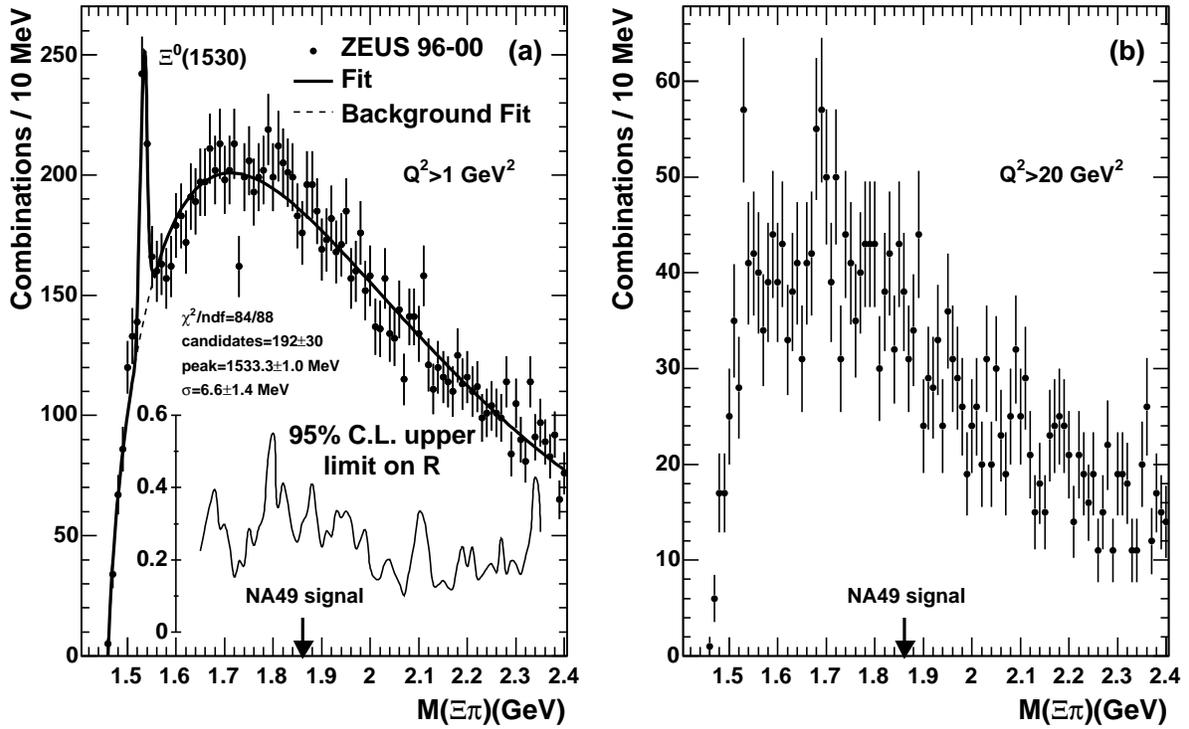}%
\caption{The $\Xi\pi$ invariant-mass spectrum for: (a) $Q^{2}>1\gev^{2}$ 
and, (b) $Q^{2}>20\gev^{2}$ (all four charge combinations summed).
The solid line in (a) is the result of a fit to the data using  
a Gaussian plus a 
three-parameter background defined by Eq.~(\ref{bgeqn}).  
The dashed line shows the background according to this fit. 
The $95\%$ C.L. upper limit on $R$ (the ratio of the 
$\Xi^{--}_{3/2}$($\Xi^{0}_{3/2}$) signal to  
$\Xi^{0}(1530)$ as defined in the text) is also shown as a function
of the invariant mass for
$Q^{2}>1\gev^{2}$. The arrows show the location of the signal observed 
by NA49.}
\label{penta_mass_spec}
\end{center}
\end{figure}

\newpage
\begin{figure}
\begin{center}
\includegraphics[width=15.0cm]{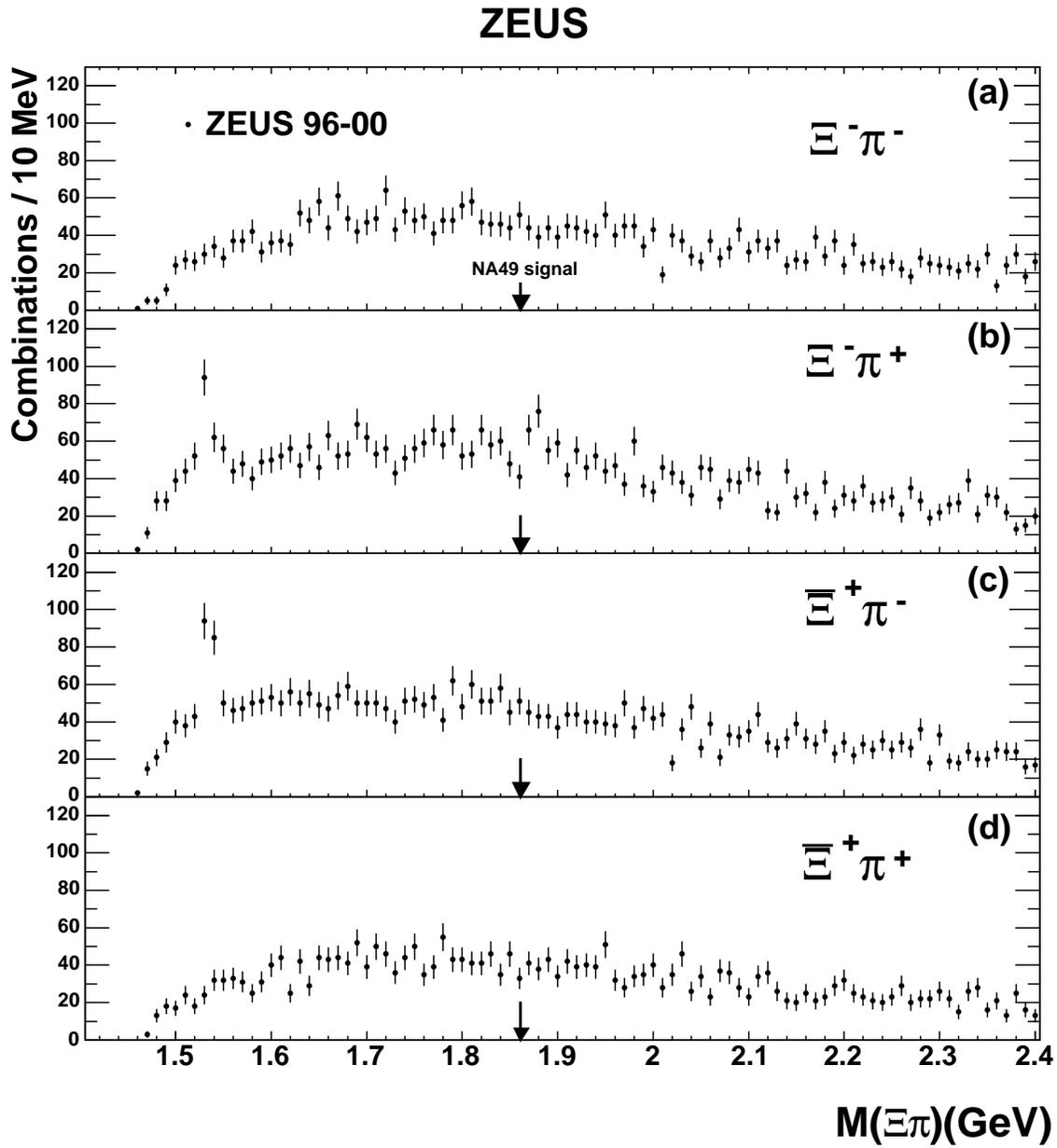}%
\caption{The $\Xi\pi$ invariant-mass spectra for each charge combination 
reconstructed at $Q^{2}>1\gev^{2}$. 
The arrows show the location of the signal observed by NA49.} 
\label{penta_mass_spec_each}
\end{center}
\end{figure}